\documentstyle[11pt]{article}

\begin{document}

\setcounter{page}{1}

\title{Unconstrained Hamiltonian formulation of low energy $SU(3)$ Yang-Mills quantum theory}
\author{H.-P. Pavel \\
        Institut f\"ur Kernphysik, c/o Theorie,
        Schlossgartenstr. 9,
        D-64289 Darmstadt, Germany \\
        and \\
        Bogoliubov Laboratory of Theoretical Physics,
        Joint Institute for Nuclear Research, Dubna, Russia\\
        }
\date{Mai 10, 2012}
\maketitle
\begin{abstract}
An unconstrained Hamiltonian formulation of the $SU(3)$ Yang-Mills quantum mechanics of spatially constant
fields is given using the method of minimal embedding of $SU(2)$ into $SU(3)$ by Kihlberg and Marnelius.
Using a canonical transformation of the gluon fields to a new set of adapted coordinates (a non-standard type polar decomposition),
which Abelianizes the Non-Abelian Gauss law constraints to be implemented, the corresponding unconstrained
Hamiltonian and total angular momentum are derived. 
This reduces the colored spin-1 gluons to unconstrained colorless
spin-0, spin-1, spin-2 and spin-3 glueball fields.
The obtained unconstrained Hamiltonian is then rewritten into a
form, which separates the rotational from the scalar degrees of freedom.
It is shown that the chromomagnetic potential has classical zero-energy valleys
for two arbitrarily large classical glueball fields, which are the unconstrained analogs 
of the well-known "constant Abelian fields". On the quantum level, practically 
all glueball excitation energy is expected to go into the increase of the strengths of these two fields.
Finally, as an outlook, the straightforward generalization to low energy $SU(3)$ Yang-Mills quantum theory
in analogy to the $SU(2)$ case is indicated, leading to an expansion in the number of spatial derivatives, equivalent to a strong coupling expansion,
with the $SU(3)$ Yang-Mills quantum mechanics constituting the leading order.
\end{abstract}


\section{\large\bf Introduction}
For a complete and detailed understanding of the low-energy hadronic properties from QCD,
such as color confinement, chiral symmetry breaking,
the formation of condensates and flux-tubes, and the spectra and strong interactions of hadrons,
it might be advantageous to first reformulate QCD in terms of gauge invariant dynamical variables,
before any approximation schemes are applied (see e.g.\cite{Christ and Lee}).
Using a canonical transformation of the dynamical variables,
which Abelianizes the Non-Abelian Gauss-law constraints,
such a reformulation has been acchieved for pure $SU(2)$ Yang-Mills theory
on the classical \cite{GKMP,KP1,KMPR} and on the quantum level \cite{pavel2}.
The resulting unconstrained
$SU(2)$ Yang-Mills Hamiltonian admits a systematic strong-coupling expansion in powers of $\lambda=g^{-2/3}$,
equivalent to an expansion in the number of spatial derivatives.
The leading order term in this expansion constitutes
the unconstrained Hamiltonian of $SU(2)$ Yang-Mills quantum mechanics of spatially constant gluon fields
\cite{Luescher}-\cite{pavel1}, for which the low-energy spectra can be calculated with high accuracy.
In recent work \cite{pavel3} is its generalization to the case
of $SU(2)$ Dirac-Yang-Mills quantum mechanics of quark and gluon fields has been carried out.
Subject of the present work is its generalization to the case of $SU(3)$.

The so-called Yang-Mills mechanics originates from Yang-Mills field theory
under the supposition  of the spatial homogeneity of the gauge  fields.
In this case  the Lagrangian of pure $SU(3)$ Yang-Mills theory
reduces to
\footnote{ Everywhere in the paper we put the spatial volume $V= 1$.
 As result the coupling constant $g$ becomes dimensionful
 with $g^{2/3}$ having the dimension of energy. The volume dependence
 can be restored in the final results by replacing $g^2$ with $g^2/V$. }
\begin{equation} \label{hl}
L={1\over 2}\left(\dot{A}_{ai}-g\ f_{abc}A_{b0} A_{ci}\right)^2
   -{1\over 2} B_{ai}^2(A) ~,
\end{equation}
with  the magnetic field
$B_{ai}(A)= (g/2) f_{abc}\epsilon_{ijk}A_{bj}A_{ck}$.
The local $SU(3)$ gauge invariance and the rotational invariance
of the original Yang-Mills action
reduces to the symmetry under the adjoined transformations 
(8-parameter subgroup of the complete 28-parameter $SO(8)$) local in time
\begin{eqnarray}
 A^{\omega}_{a0}(t)\!\!\!&=&\!\!\!
O(\omega(t))_{ab}A_{b0}(t) -\frac{1}{2g}
f_{abc}\left(O(\omega(t))\dot O(\omega(t)) \right)_{bc}\,,\nonumber\\
 A^{\omega}_{ai}(t)\!\!\!&=&\!\!\!
O(\omega(t))_{ab}A_{bi}(t)~,
\label{tr}
\end{eqnarray}
 and the global spatial  $SO(3)$  rotations
$A^{\chi}_{ai}=R(\chi)_{ij} A_{aj}$.

The canonical Hamiltonian obtained from (\ref{hl}) via Legendre-transformation reads
\begin{equation}
H_C={1\over 2}\Pi_{ai}\Pi_{ai}+{1\over 2} B_{ai}^2(A)
   + g A_{a0} \left(f_{abc} A_{bi}\Pi_{ci}\right)~,
\end{equation}
where $\Pi_{ai}$ are the momenta canonical conjugate to the spatial components $A_{ai}$.
In the constrained Hamiltonian formulation (see e.g.\cite{Christ and Lee})
the time dependence of the gauge transformations (\ref{tr}) is exploited
to put the Weyl gauge $A_{a0} = 0~$, a=1,..,8 and
the physical states $\Psi$ have to satisfy both the Schr\"odinger equation
and the three Gauss law constraints
\begin{eqnarray}
H\Psi &=& {1\over 2}
\sum_{a,i} \left[\left(\frac{\partial}{\partial A_{ai}}\right)^2+B_{ai}^2(A)\right]\Psi=E\Psi~,\\
G_a\Psi &=& -iß f_{abc}A_{bi}\frac{\partial}{\partial A_{ci}}\Psi=0~,\quad a=1,...,8~.\label{G_a}
\end{eqnarray}
The $G_a$ are the generators of the residual time independent gauge transformations,
satisfying $[G_a,H]=0$ and $[G_a,G_b]=i\epsilon_{abc}G_c$.
Furthermore $H$ commutes with the angular momentum operators
\begin{equation}
J_i  =  -i\epsilon_{ijk}A_{aj}\partial/\partial A_{ak}~,\quad i=1,2,3.
\end{equation}
The matrix element of an operator $O$ is given in the Cartesian form
\begin{equation}
\langle \Psi'| O|\Psi\rangle\
\propto
\int dA\
\Psi'^*(A) O \Psi(A)~.
\end{equation}
In the context of a weak coupling expansion, the lowest energy eigenstates of $SU(3)$ quantum mechanics
have been obtained in \cite{Weisz} systematically constructing a gauge invariant basis of low energy trial states in a variational approach.
In order to find its eigenstates in an effective way with high accuracy at least for the lowest states,
also in view of the above mentioned strong coupling expansion,
it is desirable to have a corresponding unconstrained Schr\"odinger equation.
First steps on the classical level have been done in \cite{Dahmen and Raabe} (and on the light-cone in \cite{Gerdt2}).
The basic ideas of such an unconstrained  approach
to $SU(3)$ Yang-Mills quantum mechanics will be presented in the following.
In the conclusions its straightforward generalization to low energy field-theory is indicated.

\section{\large\bf Unconstrained Hamiltonian }
\subsection{\normalsize\bf Minimal embedding of $SU(2)$ into $SU(3)$}

We follow the idea of Kihlberg and Marnelius \cite{ Kihlberg and Marnelius}
and define generators $\{\tau_a\}$ of the Lie algebra $su(3)$ such that the first three basis elements
represent the so-called {\it minimal embedding} of the subalgebra $su(2)$ in $su(3)$ (see e.g. \cite{Cornwell,Olive,Dahmen and Raabe})
\begin{eqnarray}
\label{minimal embedding}
&&\!\!\!\!\!\!\!\!\!\!\!\!
\tau_1 := \lambda_7 = \left(\begin{array}{c c c} 0&0&0\\ 0&0&-i\\ 0&i&0  \end{array}\right)
\quad
\tau_2 := -\lambda_5 =  \left(\begin{array}{c c c} 0&0&i\\ 0&0&0\\ -i&0&0  \end{array}\right)
\quad
\tau_3 := \lambda_2 =  \left(\begin{array}{c c c} 0&-i&0\\ i&0&0\\ 0&0&0  \end{array}\right)
\nonumber\\
&&\!\!\!\!\!\!\!\!\!\!\!\!
\tau_4 := \lambda_6 =  \left(\begin{array}{c c c} 0&0&0\\ 0&0&1\\ 0&1&0  \end{array}\right)
\quad\quad\!\!
\tau_5 := \lambda_4 =  \left(\begin{array}{c c c} 0&0&1\\ 0&0&0\\ 1&0&0  \end{array}\right)
\quad\quad\
\tau_6 := \lambda_1 =  \left(\begin{array}{c c c} 0&1&0\\ 1&0&0\\ 0&0&0  \end{array}\right)
\nonumber\\
&&\!\!\!\!\!\!\!\!\!\!\!\!
\tau_7 := \lambda_3 =  \left(\begin{array}{c c c} 1&0&0\\ 0&-1&0\\ 0&0&0  \end{array}\right)
\quad
\tau_8 := \lambda_8 = {1\over \sqrt{3}} \left(\begin{array}{c c c} 1&0&0\\ 0&1&0\\ 0&0&-2  \end{array}\right)~,
\end{eqnarray}
where $\lambda_a$ are the well-known Gell-Mann matrices. 
The antisymmetric matrices $(\tau_1,\tau_2,\tau_3)$ form the cyclic triplet of spin-1 matrices,
 and the traceless symmetric matrices $(\tau_4,\tau_5,\tau_6,\tau_7,\tau_8)$
represent a basis of spin-2 matrices consisting of a cyclic triplet $(\tau_4,\tau_5,\tau_6)$ and the diagonal doublet $(\tau_7,\tau_8)$.
The corresponding non-trivial non-vanishing structure constants $c_{abc}$
\begin{equation}
[{\tau_a\over 2},{\tau_b\over 2}]=  i c_{abc} {\tau_c\over 2}~,
\end{equation}
are listed in Tab. 2. 

\begin{table}[b]
\caption{Non-vanishing structure constants $c_{abc}$ w.r.t. the basis $\{\tau_a\}$.}

\begin{center}
$
\begin{array}{|c|c|c||c|}
\hline
\quad\quad a \quad\quad &\quad\quad b \quad\quad &\quad\quad c \quad\quad &\quad\quad c_{abc} \quad\quad\\
\hline\hline
   1  &  2   &  3  &   {1\over 2}  \\ \hline
   1  &   4   &  7   &    {1\over 2}  \\
   1  &   4   &  8   &   -{1\over 2}\sqrt{3}   \\
   1  &   5   &  6   &    -{1\over 2}  \\ \hline
   2  &   4   &  6   &    {1\over 2}  \\
   2  &   5   &  7   &    {1\over 2}  \\
   2  &   5   &  8   &     {1\over 2}\sqrt{3} \\ \hline
   3  &   4   &  5   &    -{1\over 2}  \\
   3  &   6   &  7   &    -1  \\
\hline
\end{array}
$
\end{center}
\end{table}

\subsection{Cartan decomposition and Euler representation of $SU(3)$}

In order to construct a corresponding Euler representation of the $SU(3)$ group,
we note that the choice (\ref{minimal embedding}) leads to the following Cartan decomposition\footnote{
Note that in \cite{Byrd,Tilma,Gerdt} a  Euler representation of the $SU(3)$ group is constructed,
which in difference to ours, is based on the trivial $su(2)$ subalgebra spanned by  $(\lambda_1,\lambda_2,\lambda_3)$, leading to
 the Cartan decomposition into ${\cal K}$ spanned
by $(\lambda_1,\lambda_2,\lambda_3,\lambda_8)$ and a vector space ${\cal P}$ spanned by $(\lambda_4,\lambda_5,\lambda_6,\lambda_7)$.
}
 of the $su(3)$-algebra into the direct sum of a vector space ${\cal K}$ spanned
by $(\tau_1,\tau_2,\tau_3)$ and a vector space ${\cal P}$ spanned by $(\tau_4,\tau_5,\tau_6,\tau_7,\tau_8)$,
\begin{equation}
su(3) = {\cal K}\oplus {\cal P}~,
\end{equation}
such that
\begin{equation}
[ {\cal K}, {\cal K}]\subset  {\cal K}~,\quad [ {\cal K}, {\cal P}]\subset  {\cal P}~,\quad [ {\cal P}, {\cal P}]\subset  {\cal K}~.
\end{equation}
This induces the corresponding Cartan decomposition on the group $SU(3)$
\begin{equation}
[SU(3)]=[SU(2)]{\rm exp}( {\cal P})~.
\end{equation}
The second factor ${\rm exp}( {\cal P})$ can be represented as a product of one parameter subgroups by sandwiching
an element between two different copies of $SU(2)$,
\begin{eqnarray}
{\rm exp}( {\cal P})=[SU(2)^\prime]\ {\rm exp}( -i(\theta \tau_7+\phi\tau_8))\ [SU(2)^{\prime\prime}]~,
\end{eqnarray}
arriving at 
\begin{equation}
[SU(3)]=[SU(2)^\prime]\ {\rm exp} ( -i\theta \tau_7) {\rm exp}(-i\phi\tau_8)\ [SU(2)]~.
\end{equation}
By choosing the Euler representation of an element of $SU(2)$,
\begin{equation}
\label{Rotfun}
R (\alpha,\beta,\gamma)=\exp(-i\alpha \tau_3)\exp(-i\beta \tau_1)\exp(-i\gamma  \tau_3)~,
\end{equation} 
we obtain the generalized Euler representation of an element $U \in SU(3) $
\begin{equation}
U=R (\alpha^\prime,\beta^\prime,\gamma^\prime)
\ {\rm exp} ( -i\phi \tau_7) {\rm exp}(-i\eta\tau_8)\  
R (\alpha,\beta,\gamma)~.
\end{equation}
For the adjoined representation $O=Ad(U)$ we therefore obtain the Euler representation
\begin{equation}
\label{adjoint SU3}
O(\alpha^\prime,\beta^\prime,\gamma^\prime,\phi,\eta,\alpha,\beta,\gamma)=R^{\rm (ad)}(\alpha^\prime,\beta^\prime,\gamma^\prime)\ O^\prime(\phi,\eta)\  R^{\rm (ad)}(\alpha,\beta,\gamma)~,
\end{equation}
with the $8\times 8$ matrices 
\begin{equation}
\label{Rotad}
R^{\rm (ad)}(\alpha,\beta,\gamma)=
\exp\left(-i\alpha Ad(\tau_3)\right)\exp\left(-i\beta Ad(\tau_1)\right)\exp\left(-i\gamma Ad(\tau_3)\right)=
\left( \begin{array}{c | c }
 R (\alpha,\beta,\gamma) & 0 \\ \hline 0 & D^{(2)}(\alpha,\beta,\gamma)  
 \end{array}\right),
\end{equation}
representing the adjoint of $R (\alpha,\beta,\gamma)$ in (\ref{Rotfun}), and 
\begin{equation}
O^\prime(\phi,\eta):=\left(\begin{array}{c | c c}
{\rm diag}\left(\cos(\phi-\eta),\cos(\phi+\eta),\cos(2\phi)\right)&
{\rm diag}\left(-\sin(\phi-\eta),-\sin(\phi+\eta),\sin(2\phi)\right)
&0
\\ \hline
{\rm diag}\left(\sin(\phi-\eta),\sin(\phi+\eta),-\sin(2\phi)\right) &
{\rm diag}\left(\cos(\phi-\eta),\cos(\phi+\eta),\cos(2\phi)\right)
&0\\0 &0&1_2
\end{array}\right)~.
\end{equation}

\subsection{Matrix decomposition into "symmetric" and "antisymmetric" parts}

Since no element of $su(3)$ commutes with all three $(\tau_1,\tau_2,\tau_3)$, the $8\times (8 \times 3) $ matrix $ c_{bai}$
 of the $SU(3)$ structure constants has maximum rank  and one can use it to decompose any rectangle $8\times 3$ matrix $A_{ai}$ into 
a "symmetric" and an "antisymmetric" part
\begin{equation}
\label{decomposition}
A_{ai}=\widehat{A}_{ai}+\sum_{b=1}^8 c_{aib} A_b~,\quad\quad  \sum_{a=1}^8\sum_{i=1}^3 c_{bai} \widehat{A}_{ai}=0~,\quad b=1,...,8~,
\end{equation}
with the "symmetric" part
\begin{eqnarray}
\label{Asym}
\widehat{A}_{ai}
&=&
\left(\begin{array}{c c c}
A_{11}                  & {1\over \sqrt{2}} {\cal A}_{3} & {1\over \sqrt{2}} {\cal A}_{2} \\
{1\over \sqrt{2}} {\cal A}_{3} & A_{22}                   & {1\over \sqrt{2}}  {\cal A}_{1} \\
{1\over \sqrt{2}} {\cal A}_{2} & {1\over \sqrt{2}} {\cal A}_{1} &  A_{33}
\\ \hline
{1\over \sqrt{3}}A_{W_0} &
{1\over \sqrt{2}}A_{X_3} +{1\over \sqrt{3}}A_{W_3}&
{1\over \sqrt{2}}A_{X_2} -{1\over \sqrt{3}}A_{W_2}
\\
{1\over \sqrt{2}}A_{X_3} -{1\over \sqrt{3}}A_{W_3}&
{1\over \sqrt{3}}A_{W_0} &
{1\over \sqrt{2}}A_{X_1} +{1\over \sqrt{3}}A_{W_1}
\\
{1\over \sqrt{2}}A_{X_2} +{1\over \sqrt{3}}A_{W_2}&
{1\over \sqrt{2}}A_{X_1} -{1\over \sqrt{3}}A_{W_1}&
{1\over \sqrt{3}}A_{W_0}
\\  
-{\sqrt{3}\over 2}A_{Y_1} +{1\over 2\sqrt{3}}A_{W_1}&
{\sqrt{3}\over 2}A_{Y_2}  +{1\over 2\sqrt{3}}A_{W_2}&
-{1\over \sqrt{3}}A_{W_3}
\\
-{1\over 2}A_{Y_1} -{1\over 2}A_{W_1}&
-{1\over 2}A_{Y_2} +{1\over 2}A_{W_2}&
A_{Y_3}
\end{array}\right)~,
\end{eqnarray}
in terms of the 5 cyclic triplets $(A_{ii})$,$({\cal A}_i)$,$(A_{X_i})$,$(A_{Y_i})$, and $(A_{W_i})$, (i=1,2,3) and the singlet $A_{W_0}$,
and the "antisymmetric" part
\begin{eqnarray}
\sum_{b=1}^8 c_{aib} A_b
&=&
\left(\begin{array}{c c c}
0         & A_{3}/2 & -A_{2}/2 \\
-A_{3}/2 & 0        & A_{1}/2 \\
A_{2}/2 & -A_{1}/2  &  0
\\ \hline
-(A_7- \sqrt{3}A_8)/2 & -A_6/2 & A_5/2\\
A_6/2 & -(A_7+ \sqrt{3}A_8)/2  & -A_4/2
\\
-A_5/2 & A_4/2 & A_7
\\  
A_4/2 & A_5/2 & -A_6
\\
- \sqrt{3}A_4/2 & \sqrt{3} A_5/2 & 0
\end{array}\right)~,
\end{eqnarray}
in terms of the triplets $(A_{1},A_{2},A_{3})$ and $(A_{4},A_{5},A_{6})$ and the doublet $(A_{7},A_{8})$. 

\begin{equation}
\label{absolut}
 \sum_{a=1}^8\sum_{i=1}^3  \left( A_{ai}\right)^2 
= \left[\left(A_{11}^2+{\cal A}_1^2+A_{X_1}^2+A_{Y_1}^2+A_{W_1}^2\right)+{\rm cycl. perm.}+A_{W_0}^2\right]
+ \left[{1\over 2}\sum_{a=1}^3 A_{a}^2+{3\over 2}\sum_{a=4}^8 A_{a}^2\right]~.
\end{equation}
The representation of $\widehat{A}$ chosen in (\ref{Asym}) is the unique one, which satisfies both the "symmetry" condition
$c_{abi} \widehat{A}_{bi}=0$ in (\ref{decomposition}) and
the diagonality requirement, that the sum of the squares $\sum_{a i} \widehat{A}_{ai}^2$ 
 (the first square bracket on the right hand side of (\ref{absolut}))  is the sum of the squares of the new elements with unit coefficient.
Due to the choice of the minimal embedding (\ref{minimal embedding}) 
we have additionally symmetry under cyclic permutation of the indices 1,2,3.

\subsection{\normalsize\bf Hamiltonian reduction and the symmetric gauge for $SU(3)$}

The local symmetry transformation (\ref{tr}) of the gauge potentials
\( A_{ai} \) prompts us with the set of coordinates in terms of which the
separation of  the  gauge degrees of freedom occurs.
Carrying out the following coordinate transformation from the 24 elements of the \(8\times 3\)  matrix $A_{ai}$
to 24 new coordinates, the 8 angles $q_1,..q_8$ of an orthogonal $8\times 8$ matrix \( O (q_1,..,q_8)  \)
representing the adjoint representation of $SU(3)$, e.g. given by the Euler angles in (\ref{adjoint SU3}),
 and 16 elements collected into the "symmetric"  \(8\times 3\)  matrix $ \widehat{S}$ of the form (\ref{Asym}),  satisfying $c_{abi} \widehat{S}_{bi}=0$, via
\begin{equation}
\label{eq:pcantr}
A_{ak} \left(q, \widehat{S} \right)
= O_{a\hat{a}}\left( q_1,..,q_8 \right) \widehat{S} _{\hat{a} k}~,
\end{equation}
where the matrix $\widehat{S}$ is given by
\begin{eqnarray}
\label{Shat}
 \widehat{S}_{\hat{a} k }\equiv {S_{i k }\choose \overline{S}_{A k} }
&=&
\left(\begin{array}{c c c}
S_{11}                  & {1\over \sqrt{2}} {\cal S}_{3} & {1\over \sqrt{2}} {\cal S}_{2} \\
{1\over \sqrt{2}} {\cal S}_{3} & S_{22}                   & {1\over \sqrt{2}}  {\cal S}_{1} \\
{1\over \sqrt{2}} {\cal S}_{2} & {1\over \sqrt{2}} {\cal S}_{1} &  S_{33}
\\ \hline
{1\over \sqrt{3}}W_0 &
{1\over \sqrt{2}}X_3 +{1\over \sqrt{3}}W_{3}&
{1\over \sqrt{2}}X_2 -{1\over \sqrt{3}}W_{2}
\\
{1\over \sqrt{2}}X_3 -{1\over \sqrt{3}}W_{3}&
{1\over \sqrt{3}}W_0 &
{1\over \sqrt{2}}X_1 +{1\over \sqrt{3}}W_{1}
\\
{1\over \sqrt{2}}X_2 +{1\over \sqrt{3}}W_{2}&
{1\over \sqrt{2}}X_1 -{1\over \sqrt{3}}W_{1}&
{1\over \sqrt{3}}W_0
\\  
-{\sqrt{3}\over 2}Y_1 +{1\over 2\sqrt{3}}W_{1}&
{\sqrt{3}\over 2}Y_2  +{1\over 2\sqrt{3}}W_{2}&
-{1\over \sqrt{3}}W_{3}
\\
-{1\over 2}Y_1 -{1\over 2}W_{1}&
-{1\over 2}Y_2 +{1\over 2}W_{2}&
Y_3
\end{array}\right)
\end{eqnarray}
with the positive definite, symmetric \(3\times 3\)  matrix \( S \), and the $5\times 3$ matrix $\overline{S}$ which is a function of 
ten fields, the triplets $(X_1,X_2,X_3)$,  $(Y_1,Y_2,Y_3)$, and $(W_1,W_2,W_3)$ and the singlet $W_0$.
Note that, as for the $SU(2)$ case,
 the symmetric tensor field $S$ can be decomposed
 into the spin-0 and  spin-2 components\footnote{
We use the Hermitean combinations 
$S_{|M|+}^{(J)}:=-i^{|M|} (S_{+|M|}^{(J)}+ S_{-|M|}^{(J)})/ \sqrt{2} $ and
$S_{|M|-}^{(J)}:=-i^{|M|+1} (S_{+|M|}^{(J)}- S_{-|M|}^{(J)})/ \sqrt{2} $ for $|M|>0$.
}
\begin{equation}
S_{ik}=C_{1i\ 1k}^{2 A}\ S^{\!(2)}_{A}+\frac{1}{\sqrt{3}}\delta_{ik}\ S^{\!(0)}
\end{equation}
with
\begin{eqnarray}
 S^{\!(0)}&=&\frac{1}{\sqrt{3}}(S_{11}+S_{22}+S_{33})
\nonumber\\
 S^{\!(2)}=(S^{\!(2)}_{1+},S^{\!(2)}_{1-},S^{\!(2)}_{2-},S^{\!(2)}_{2+},S^{\!(2)}_0)
&=&\left({\cal S}_1,{\cal S}_2,{\cal S}_3,{1\over\sqrt{2}}(S_{11}-S_{22}),
\sqrt{2\over 3}(S_{33}-{1\over 2}S_{11}-{1\over 2}S_{22})\right)
\label{SSpin02}
\end{eqnarray}

\noindent
In a similar way, the $5\times 3$ matrix $\overline{S}$ can be decomposed into spin-1 and spin-3 fields,
\begin{eqnarray}
\label{xy}
 {X_{i}\choose Y_{i} }
 \equiv \left(
       \begin{array}{c c}\sqrt{3/5}& \sqrt{2/5}\cr -\sqrt{2/5} &\sqrt{3/5}\end{array}\right)
                              {V^{(1)}_{i}\choose \overline{W}_{i} }~,\quad i=1,2,3~,
\end{eqnarray}
with the cyclic triplet of 3  Cartesian components
$V^{(1)}_i,i=1,2,3$, of a vector-field and the 7  Hermitian components  $W_M^{(3)}$ of a spin-3 field, 
written in the form of a cylic singlet $W_0:=W_{2-}^{(3)}$, and the cyclic triplet combinations
\begin{eqnarray}
\label{w}
&&\!\!\!\!\!\!\!\!\!\!\!\!\!\!\!\!\!\!\!\!\!\!\!\!\!\!\!\!\!\!\!\!\!\!\!\!
{ \overline{W}_1\choose W_{1}}
 \equiv \left(
       \begin{array}{c c}\sqrt{3/8}& -\sqrt{5/8}\cr \sqrt{5/8} &\sqrt{3/8}\end{array}\right)
                              {W_{1-}^{(3)}\choose W_{3-}^{(3)} }~,\quad
{  \overline{W}_2\choose W_{2}} 
 \equiv  \left(
       \begin{array}{c c}\sqrt{3/8}& \sqrt{5/8}\cr -\sqrt{5/8} &\sqrt{3/8}\end{array}\right)
                            {W_{1+}^{(3)}\choose W_{ 3+}^{(3)} }~,\quad
 { \overline{W}_3\choose W_{3}}
 \equiv {W_{0}^{(3)}\choose W_{2+}^{(3)}}~.
\end{eqnarray}

\noindent
The decomposition
(\ref{eq:pcantr}) can be seen as imposing the "symmetric gauge" 
\begin{eqnarray}
\label{symgauge}
\chi_a(A)=\sum_{b=1}^8\sum_{i=1}^3 c_{abi} A_{bi}=0~,\quad a=1,...,8~.
\end{eqnarray}
For an investigation of the existence and uniqueness of the non-standard polar decomposition (\ref{eq:pcantr}), 
 let us consider the 6 components of the symmetric 
$s_{ik}:=A_{ai}A_{ak}=\widehat{S}_{\hat{a}i}\widehat{S}_{\hat{a}k}=S^2_{ik}+\overline{S}^T_{iA}\overline{S}_{Ak}$  
and the 10 components of the totally symmetric    $\overline{s}_{ijk}:=d_{abc}A_{ai}A_{bj}A_{ck}=d_{\hat{a}\hat{b}\hat{c}}\widehat{S}_{\hat{a}i}\widehat{S}_{\hat{b}j}\widehat{S}_{\hat{c}k}
=d_{lmA}S_{li}S_{mj}\overline{S}_{Ak}+d_{ABC}\overline{S}_{Ai}\overline{S}_{Bj}\overline{S}_{Ck}$, with the totally symmetric $SU(3)$ coefficients $d_{abc}$ written in the minimal embedding basis (\ref{minimal embedding}). It follows directly that the $3\times 3$
submatrix $S$ can be chosen positive definite, as for the $SU(2)$ case. The algebraic task, to prove the existence and uniqueness of $\widehat{S}$, by
expressing it in terms of $s$ and $\overline{s}$, goes beyond the scope of this work.
We only point out here, that the totally symmetric $3\times 3\times 3$ matrix $\overline{s}$ can be written in the form of a $5\times3$ matrix  $\overline{S}^\prime$ as in (\ref{Shat}), 
with $X_1^\prime:= {1\over \sqrt{2}}(s_{122}+s_{133})$,
 $Y_1^\prime:=s_{111} $, $W_1^\prime:= {1\over \sqrt{2}}(-s_{122}+s_{133})$, and their cyclic permutations, and $W_0^\prime:=s_{123} $.

\noindent
The corresponding momenta $\Pi_{ai}$ are found in terms of the new  coordinates $ (q_{\hat{a}}, \widehat{S}_\alpha)$ and 
momenta  $(p_{\hat{a}}, \widehat{P}_\alpha)$ as
\begin{eqnarray}
\Pi_{ak}= O_{a \hat{a}}\left( q \right)\left( \widehat{P}_{\hat{a}k}
-c_{\hat{a}k\hat{b}}\gamma^{-1}_{\hat{b}\hat{c}}\left(T_{\hat{c}}-\Omega^{-1}_{\hat{c}\hat{d}}(q)p_{\hat{d}}\right)\right)~,
\end{eqnarray}
where the $8\times 3$ matrix $ \widehat{P}$  is of the same form as $ \widehat{S}$ with the elements 
$\widehat{S}_\alpha:=(\{S_{ii}\},\{{\cal S}_i\},\{X_i\},\{Y_i\},\{W_i\},W_0)$
replaced with the canonical momenta $\widehat{P}_\alpha:=-i\partial/\partial\widehat{S}_\alpha$. The Faddeev-Popov(FP) matrix 
 $\gamma$ and the operators $T$ are defined as
\begin{eqnarray}
\label{FP}
\gamma_{\hat{a}\hat{b}}&:=&-c_{\hat{a}\hat{c}i}c_{\hat{b}\hat{c}\hat{d}} \widehat{S}_{\hat{d}i}~,
\\
\label{T}
T_{\hat{a}}&:=&c_{\hat{a}\hat{b}\hat{c}} \widehat{S}_{\hat{b}i}\widehat{P}_{\hat{c}i}~,  
\end{eqnarray}
and  $\Omega_{ab}(q) := (1/2) \mbox{tr}
\left[\left(U^{-1} \partial U/\partial q_a\right)\tau_b\right]$, where $U$ in the fundamental representation is related to $O$ via
$O_{ab}(q)=  (1/2) \mbox{tr}\left[U(q)\tau_b U^{-1}(q)\tau_b\right]$.
Explicitly, $\gamma$ is
\begin{equation}
\label{uncFP}
\gamma(\widehat{S})={1\over 4}
\left(\begin{array}{c | c }
S-1_3\mbox{ tr} S                  &2\overline{S}[-{3\over 2}V,W]^{T} \\
 \hline
 2\overline{S}[-{3\over 2}V,W]                 &\begin{array}{c | c }
                                                                   3 \left( S -{7\over 9}1_3 \mbox{ tr} S\right) & \begin{array}{c  c }
                                                                    -3 {\cal S}_{1}/\sqrt{2} &-\sqrt{3/2} {\cal S}_{1} \\ 
                                                               3  {\cal S}_{2}/\sqrt{2} &-\sqrt{3/2} {\cal S}_{2} \\ 0 & \sqrt{6}  {\cal S}_{3}
                                                                          \end{array}\\ \hline \begin{array}{c c c }
                                                                    -3  {\cal S}_{1}/\sqrt{2} & 3  {\cal S}_{2}/\sqrt{2} & 0 \\  
                                                             -\sqrt{3/2} {\cal S}_{1} &-\sqrt{3/2} {\cal S}_{2} & \sqrt{6}  {\cal S}_{3}
                                                                          \end{array} & \begin{array}{c  c }
                                                                    -( S_{11}+S_{22}+4S_{33}) &\sqrt{3}( S_{11}-S_{22}) \\ 
                                                              \sqrt{3} ( S_{11}-S_{22}) & -3( S_{11}+S_{22})  
                                                                          \end{array}
                                                                          \end{array} 
\end{array}\right).
\end{equation}
The magnetic field finally becomes
\begin{eqnarray}
\label{uncB}
B_{ak}(A)= O_{a\hat{a}}\left( q \right)\left( \widehat{B}_{\hat{a}k}(\widehat{S})+c_{\hat{a}k\hat{b}}B_{\hat{b}}(\widehat{S})\right)~,
\end{eqnarray}
where the $8\times 3$ matrix $ \widehat{B}$  is of the same form as $ \widehat{S}$ with the elements 
$\widehat{S}_\alpha$ replaced by the corresponding $\widehat{B}_\alpha$, and the
antisymmetric components are $B_{\hat{a}}\equiv(B^{(1)}_i,B^{(2)}_A)$, with the vector part $B^{(1)}_i=0$ as for the $SU(2)$ case.

The Jacobian of the transformation (\ref{eq:pcantr}) is the product of the Haar measure $|\Omega|$ and the FP determinant $|\gamma|$
\begin{equation}
|\partial A/\partial(q,\widehat{S})| \propto
|\Omega|\  |\gamma|~.
\end{equation}
The range of the variables has to be chosen such that $\gamma(\widehat{S})$ is invertible and hence the transformation (\ref{eq:pcantr}) well-defined.

\subsection{Unconstrained Hamiltonian and angular momentum}

The variables \(  \widehat{S}_\alpha \) make no contribution to the
Gauss law operators
$G_a = -iO_{a\hat{a}}(q) \Omega^{-1}_{\ \hat{b}\hat{c}}(q)\partial/\partial q_{\hat{c}}$.
Hence, assuming the invertibility of  the matrix
$ \Omega$, the non-Abelian Gauss laws (\ref{G_a})  are
equivalent to  the set of Abelian  constraints
\begin{equation}
G_a \Phi  = 0  \quad \forall\    a=1,...,8      
 \quad \quad    
\Leftrightarrow    
\quad \quad    
 \partial \Phi/\partial q_{\hat{a}}  = 0  \quad  \forall\ \hat{a}=1,...,8.  
\quad \quad  ({\rm Abelianization})
\end{equation}
and the physical  Hamiltonian of $SU(3)$ Yang-Mills quantum mechanics reads
\begin{eqnarray}
\label{uncYMH}
\!\!\!\!\!\!\!\! H\!\!\!\! &=&\!\!\!\! - {1\over 2}|\gamma|^{-1}\!\sum_{\alpha}\!\!
{\partial\over \partial \widehat{S}_{\alpha}}|\gamma|{\partial\over \partial \widehat{S}_{\alpha}}
+{1\over 4}|\gamma|^{-1}\!\!\sum_{\hat{a},\hat{b}=1}^8\!\! 
T_{\hat{a}} |\gamma|\left(\gamma^{-1}_{\hat{a}i}\gamma^{-1}_{i \hat{b}}+3\gamma^{-1}_{\hat{a}A}\gamma^{-1}_{A \hat{b}}\right) T_{\hat{b}}  
+{1\over 2}\left(\sum_{\alpha}\widehat{ B}^2_{\alpha}
+{3\over 2}\sum_{A} B_{A}^{(2)2}\right),
\end{eqnarray}
and the matrix element of a physical operator O is given by
\begin{equation}
\label{uncME}
\langle \Psi'| O|\Psi\rangle\
\propto
\int \prod_\alpha d \widehat{S}_\alpha\ 
|\gamma |\
\Psi'^*( \widehat{S})\ O\ \Psi( \widehat{S})~.
\end{equation}
The dependence on the pure gauge degrees of freedom $q$ has completely disappeared from both the Hamiltonian (\ref{uncYMH}) 
and the matrix elements (\ref{uncME}).
Furthermore, the physical angular momenta are obtained as
\begin{eqnarray}
\label{uncJ}
J_i=\epsilon_{ijk}\widehat{S}_{\hat{a} j}\left[\widehat{P}_{\hat{a}k} -c_{\hat{a}k\hat{c}}(\gamma^{-1}_{\hat{c}\hat{b}}T_{\hat{b}})\right]=
\epsilon_{ijk}\widehat{S}_{\hat{a}j}\widehat{P}_{\hat{a}k} +2 T_i =J_i^{(V)}+J_i^{(S)}+J_i^{(W)}~,\quad i=1,2,3~,
\end{eqnarray}
with the spin-1,spin-2, and spin-3 parts
\begin{eqnarray}
J_i^{(V)} = -i\epsilon_{ijk}V^{(1)}_{j}{\partial \over\partial V^{(1)}_{k}}~, \quad \quad
J_i^{(S)} = -2i\epsilon_{ijk}S_{lj}{\partial \over\partial S_{lk}}~,\quad \quad
J_i^{(W)} = -2\sqrt{7}i C^{1 i}_{3 M\ 3N}W^{(3)}_{M}{\partial \over\partial W^{(3)}_{N}}~, \quad \quad
\end{eqnarray}
in terms of the physical variables.
Hence the physical fields $S^{(0)}$, $V^{(1)}_i$, $S^{(2)}_A$, and
$W^{(3)}_M$, indeed transform as spin-0, spin-1,spin-2, and spin-3 fields under spatial rotations
\begin{eqnarray}
\label{traforot1}
S^{(0)\prime}=S^{(0)}~,
\quad\quad\quad
V^{(1)\prime}_{i^\prime} = R(\chi)_{i^\prime i}V^{(1)}_i~,
\quad\quad\quad
S^{(2)\prime}_{A^\prime}= D^{(2)}_{A^\prime A}(\chi)S^{(2)}_A~,
\quad\quad\quad
W^{(3)\prime}_{M^\prime}= D^{(3)}_{M^\prime M}(\chi)\ W^{(3)}_{M}~,
\end{eqnarray}
which is equivalent to the transformation
\begin{equation}
\label{traforot}
\widehat{S}^\prime_{\hat{a}^\prime k^\prime}\equiv
{S^\prime_{i^\prime k^\prime}\choose \overline{S}^\prime_{A^\prime k^\prime} }=
\left(\!\!\! \begin{array}{c  c }
 R_{i^\prime i}(\chi) & 0 \\  0 & D^{(2)}_{A^\prime A}(\chi)  
\!\!\! \end{array}\right)
{S_{i k}\choose \overline{S}_{A k} } R^{T}_{k k^\prime}(\chi)
\equiv R^{\rm (ad)}_{\hat{a}^\prime\hat{a}}(\chi)R_{k^\prime k}(\chi) \widehat{S}_{\hat{a} k}~,
\end{equation}
with the \( SO(3)\) rotation matrix \(R(\chi)\) of (\ref{Rotfun}) 
and the corresponding adjoined SO(8) matrix \(R^{\rm (ad)}(\chi)\) of (\ref{Rotad}), parametrized by the 3 Euler angles $\chi$.
The FP matrix $\gamma$ in (\ref{FP}) and the $T$ in (\ref{T})  are found to transform as
\begin{equation}
\label{gammaT}
\gamma^\prime_{\hat{a}^\prime\hat{b}^\prime}
=R^{\rm (ad)}_{\hat{a}^\prime\hat{a}}(\chi)  R^{\rm (ad)}_{\hat{b}^\prime\hat{b}}(\chi)\ \gamma_{\hat{a}\hat{b}}~, 
\quad\quad\quad\quad
 T^\prime_{\hat{a}^\prime}
=R^{\rm (ad)}_{\hat{a}^\prime\hat{a}}(\chi)\  T_{\hat{a}}~.
\end{equation}
The indices $i,j,k,..$ are therefore spin-1 indices and $A,B,C,..$ spin-2 indices.
The $T_{\hat{a}}=(T^{(1)}_i,T^{(2)}_A)$ read  
\begin{eqnarray}
\label{uncT}
T^{(1)}_i&=&{3\over 4} J_i^{(V)}+{1\over 4} J_i^{(S)}+{1\over 3} J_i^{(W)} =
{1\over 4} \left[J_i-2 J_i^{(V)}-{1\over 3} J_i^{(W)}\right]~,\quad i=1,2,3~,
\end{eqnarray}
and expressions for $T^{(2)}_A$ not shown here.
The magnetic fields $\widehat{B}$ and  $B^{(2)}_A$ in (\ref{uncB}) transform like  $\widehat{S}$ and $T^{(2)}_A$.

\section{\normalsize\bf Unconstrained Hamiltonian in terms of
rotational and scalar degrees of freedom}

A more transparent form for the unconstrained Yang-Mills
Hamiltonian (\ref{uncYMH}), maximally separating the rotational from the rotation
invariant degrees of freedom,
can be obtained using the transformation properties (\ref{traforot1})  of the canonical fields $S$,$V$ and $W$ under spatial rotations
generated by the unconstrained angular momentum (\ref{uncJ}).
We shall perform a principal-axes transformation of the positive definite symmetric matrix $S$, which is the upper submatrix of $\widehat{S}$, and shall
use as new coordinates the 3 Euler angles describing the orientation of the intrinsic system of $S$, and 13 rotation invariants,
the 3 eigenvalues of $S$ and the 10 spin-1 and spin-3 fields transformed to the intrinsic system of $S$.

\subsection{Transformation to angular and rotation invariant variables}
We limit ourselves in this work to the case of principle orbit configurations
of non-coinciding eigenvalues $\phi_1,\phi_2,\phi_3>0$ of the positive definite symmetric matrix $S$,
which without loss of generality can be taken as
\begin{equation}
\label{range}
0<\phi_1<\phi_2<\phi_3<\infty~,
\end{equation}
(not considering singular orbits where two or more eigenvalues coincide)
and perform a principal-axes transformation on the symmetric tensor field part S
\begin{eqnarray}
\label{patup}
S=:  R(\alpha,\beta,\gamma)\ \mbox{diag}\ ( \phi_1 , \phi_2 , \phi_3 ) \
R^{T}(\alpha,\beta,\gamma)~,
\end{eqnarray}
with the 3 Euler angles $\chi=(\alpha,\beta,\gamma)$ describing
 the orientation of the intrinsic system of the symmetric tensor field $S$.
Since $S$ is the upper $3\times 3$ part of the $8\times 8$ matrix $\widehat{S}$, which transforms under spatial rotations as (\ref{traforot}),
the principal-axes transformation (\ref{patup}) induces the transformation
\begin{equation}
\label{coordin}
\widehat{S}  (S^{(0)},S^{(2)},V^{(1)},W^{(3)}) =:
 R^{\rm (ad)}(\alpha,\beta,\gamma)\ \widehat{S}^{\rm intr}(\phi_1,\phi_2,\phi_3,v^{(1)},w^{(3)})\ R^T(\alpha,\beta,\gamma)~,
\end{equation}
that is explicitly
\begin{eqnarray}
\label{su3pat}
 {S(S^{(0)},S^{(2)})\choose  \overline{S}(V^{(1)},W^{(3)}) }
&=:&\left(\!\!\! \begin{array}{c  c }
 R(\alpha,\beta,\gamma) & 0 \\ \!\!  0 & D^{(2)}(\alpha,\beta,\gamma)  
 \end{array}\!\!\right)
{\mbox{diag}(\phi_1,\phi_2,\phi_3)\choose \overline{S}(v^{(1)},w^{(3)}) }  R^T(\alpha,\beta,\gamma)~,
\end{eqnarray}
from the original 16 unconstrained fields $(S^{(0)},S^{(2)},V^{(1)},W^{(3)})$ to the 3 Euler angles $(\alpha,\beta,\gamma)$
and the 13 rotation invariant fields $(\phi_1,\phi_2,\phi_3,v^{(1)},w^{(3)})$ defined with respect to the intrinsic system of the $3\times 3$
symmetric submatrix $S$.
The $5\times 3$ submatrix $ \overline{S}$ is the same functional of the intrinsic $(v^{(1)},w^{(3)})$ as for
the original spin-1 and spin-3 fields $(V^{(1)},W^{(3)})$,
and the spin-1 and spin-3 fields $v^{(1)}$ and $w^{(3)}$ relative to the intrinsic system of $S$, are defined by the lower line of (\ref{su3pat}),
which is equivalent to
\begin{eqnarray}
V^{(1)}_i =: R(\alpha,\beta,\gamma)_{i i^\prime}v^{(1)}_{ i^\prime}~,
\quad\quad\quad
W^{(3)}_M=: D^{(3)}_{MM^\prime}(\alpha,\beta,\gamma)\ w^{(3)}_{M^\prime}~.
\end{eqnarray}
Defining finally the intrinsic variables $(\{x_i\},\{y_i\},\{w_i\},w_0)$ to be the same functions of the intrinsic $(v^{(1)},w^{(3)})$ as the
original capital $(\{X_i\},\{Y_i\},\{W_i\},W_0)$  are defined in terms of the capital  $(V^{(1)},W^{(3)})$ in (\ref{xy}) and  (\ref{w}),
we can write the coordinate transformation (\ref{coordin}) also in the form
\begin{equation}
\label{coordinxy}
\widehat{S}_\alpha =  (\{S_{ii}\},\{{\cal S}_i\},\{X_i\},\{Y_i\},\{W_i\},W_0)\quad \longrightarrow 
\quad \chi=(\alpha,\beta,\gamma)\ \ \cup\ \ 
\widehat{S}^{\rm intr}_\alpha=(\{\phi_i\},\{x_i\},\{y_i\},\{w_i\},w_0).
\end{equation}
According to (\ref{SSpin02}) and (\ref{patup}), the unconstrained spin-0 and spin-2 gluon fields  can be written in the form
\begin{eqnarray}
S^{\!(0)} =\left(\phi_1+\phi_2+\phi_3\right)/\sqrt{3}~,
\quad\quad\quad
S^{\!(2)}_{A} =
\sqrt{\frac{2}{3}}\left[
\left(\phi_3-\frac{1}{2}\left(\phi_1+\phi_2\right)\right)D^{(2)}_{A0}(\chi)
+\frac{\sqrt{3}}{2}\left(\phi_1-\phi_2\right)
D^{(2)}_{A2+}(\chi)
\right],
\end{eqnarray}
in terms of the principle-axes variables,
and the corresponding canonically conjugate momenta are found as
\begin{eqnarray}
-i{\partial\over\partial S^{\!(0)}}\!\!\! &=&\!\!\!-i\left(
{\partial\over\partial \phi_{1}}+{\partial\over\partial \phi_{2}}+{\partial\over\partial \phi_{3}}\right)/\sqrt{3}~,
\nonumber\\
\label{new-mom}
-i{\partial\over\partial S^{\!(2)}_{A}}\!\!\! &=&\!\!\!
\sqrt{\frac{2}{3}}\left[
-i\left({\partial\over\partial \phi_{3}}-\frac{1}{2}\left({\partial\over\partial \phi_{1}}
+{\partial\over\partial \phi_{2}}\right)
\right)D^{(2)}_{A 0}(\chi)
-\frac{\sqrt{3}}{2}i\left({\partial\over\partial \phi_{1}}-{\partial\over\partial \phi_{2}}\right)
D^{(2)}_{A 2+}(\chi)\right]\nonumber\\
&&+\frac{1}{\sqrt{2}}
       \Bigg[ D^{(2)}_{A 1+}(\chi) \frac{\xi_1-\widetilde{J}^{(v)}_1-\widetilde{J}^{(w)}_1}{\phi_2 - \phi_3}
             +D^{(2)}_{A 1-}(\chi) \frac{\xi_2-\widetilde{J}^{(v)}_2-\widetilde{J}^{(w)}_2}{\phi_3 - \phi_1}
             +D^{(2)}_{A 2-}(\chi) \frac{\xi_3-\widetilde{J}^{(v)}_3-\widetilde{J}^{(w)}_3}{\phi_1 - \phi_2}
       \Bigg],
\end{eqnarray}
using the intrinsic angular momenta\footnote{
For the case of Euler angles $\chi=(\alpha,\beta,\gamma)$ we have
$
{\cal M}^{-1}=
\left(\begin{array}{ccc}
\sin\gamma & -\cos\gamma / \sin\beta  & \cos\gamma \cot\beta \cr
\cos\gamma &  \sin\gamma / \sin\beta  & -\sin\gamma \cot\beta  \cr
0&0&1
\end{array}\right)~.
\nonumber
$
}
\begin{equation}
\xi_i:=-i{\cal M}^{-1}_{ij}{\partial\over\partial \chi_{i}}~,
\quad
{\cal M}_{ij}:=-{1\over 2}\ \epsilon_{jst}\!
        \left(R^T{\partial R\over\partial \chi_{i}} \right)_{st}~,
\quad
[\xi_i,\xi_j]=-i\epsilon_{ijk}\xi_k~.
\end{equation}
The intrinsic momenta $\widetilde{J}^{(v)}_i$ and $\widetilde{J}^{(w)}_i$ of the spin-1 and spin-3 fields  $v^{(1)}$ and $w^{(3)}$, given by
\begin{equation}
J^{(V)}_i=R(\chi)_{i i^\prime}\widetilde{J}^{(v)}_{ i^\prime}\quad\quad
J^{(W)}_i=R(\chi)_{i i^\prime}\widetilde{J}^{(w)}_{ i^\prime}
\end{equation}
appear in (\ref{new-mom}) in order to ensure
$[\partial/\partial S^{\!(2)}_{A},\partial/\partial V^{(1)}_i]=[\partial/\partial S^{\!(2)}_{A},\partial/\partial W^{(3)}_M]=0$.

\noindent
Using the expressions (\ref{new-mom}), the total unconstrained angular momentum (\ref{uncJ}) takes the form
\begin{equation}
\label{uncJin}
J_i=R_{i i^\prime }(\chi)\xi_{ i^\prime}~.
\end{equation}
The Jacobian of  the coordinate transformation (\ref{su3pat}),
\begin{equation}
|\partial \widehat{S}/\partial(\chi,\phi,v^{(1)},w^{(3)})| \propto
\sin\beta \prod_{i<j}\left(\phi_i- \phi_j\right)~,
\end{equation}
is independent of the intrinsic spin-1 and spin-3 fields, and hence the same as for the $SU(2)$ case.

\subsection{Unconstrained Hamiltonian in terms of angular and rotation invariant variables}

Transforming also the Faddeev-Popov matrix $\gamma$ in (\ref{uncFP}) and the $T$ in (\ref{uncT}) into the intrinsic system in accordance with (\ref{gammaT})
\begin{equation}
\gamma_{\hat{a}\hat{b}}=R^{\rm (ad)}_{\hat{a}\hat{a}^\prime}(\chi)  R^{\rm (ad)}_{\hat{b}\hat{b}^\prime}(\chi)\ \widetilde{\gamma}_{\hat{a}^\prime\hat{b}^\prime} 
\quad\quad
 T_{\hat{a}}=R^{\rm (ad)}_{\hat{a}\hat{a}^\prime}(\chi)\widetilde{T}_{\hat{a}^\prime}
\end{equation}
we obtain the physical Hamiltonian (\ref{uncYMH}) in terms of the rotational and rotation invariant variables (\ref{coordinxy}) as
\begin{eqnarray}
\label{uncYMHin}
 H \!\!\!\! &=&\!\!\!\! - {1\over 2}
{\cal J}^{-1}\sum_{\alpha}
{\partial\over \partial \widehat{S}^{\rm intr}_{\alpha}}{\cal J}{\partial\over \partial \widehat{S}^{\rm intr}_{\alpha}}
+ {1\over 4}
|\widetilde{\gamma}|^{-1}\sum_{i,j,k}^{\rm cyclic} \frac{\xi_i-\widetilde{J}^{(v)}_i-\widetilde{J}^{(w)}_i}{\phi_j - \phi_k}
|\widetilde{\gamma}| \frac{\xi_i-\widetilde{J}^{(v)}_i-\widetilde{J}^{(w)}_i}{\phi_j - \phi_k}\nonumber\\
&&\quad\quad
+{1\over 4}{\cal J}^{-1}\!\!\sum_{\hat{a},\hat{b}=1}^8 \!\!
\widetilde{T}_{\hat{a}}\left(\widehat{S}^{\rm intr}_\alpha,{\partial\over \partial \widehat{S}^{\rm intr}_{\alpha}},\xi\right) {\cal J} \left(\widetilde{\gamma}^{-1}_{\hat{a}i}\widetilde{\gamma}^{-1}_{i \hat{b}}+3\widetilde{\gamma}^{-1}_{\hat{a}A}\widetilde{\gamma}^{-1}_{A \hat{b}}\right)
\widetilde{T}_{\hat{b}}\left(\widehat{S}^{\rm intr}_\alpha,{\partial\over \partial \widehat{S}^{\rm intr}_{\alpha}},\xi\right) 
+ V_{\rm magn}(\widehat{S}^{\rm intr}),
\end{eqnarray}
with the total Jacobian ${\cal J}:=|\widetilde{\gamma}|\prod_{i<j}(\phi_i-\phi_j)$ and the explicit form of the 8 components $\widetilde{T}_{\hat{a}}$
shown in the Appendix.
All dependence on the rotational variables $(\alpha,\beta,\gamma)$ in (\ref{uncYMHin})  is collected in the intrinsic angular momenta $\xi_i$
such that the vanishing of the commutator of (\ref{uncYMHin}) with the total spin (\ref{uncJin}), $[H,J_i]=0$, is trivially fulfilled due to $[J_i,\xi_j]=0$.

\noindent
The matrix elements of a physical operator $O$ given as
\begin{equation}
\langle\Psi'|O|\Psi\rangle\! \propto\!\!
\int\!\! d\alpha \sin\beta d\beta d\gamma\!\!\!\!\!\!\!\!\!\!\!\!\!
\int\limits_{0<\phi_1<\phi_2<\phi_3}\!\!\!\!\!\!\!\!\!\!\!\!\! \Big[\!\!\prod^{\rm cyclic}\! d\phi_i
\! \left(\phi_j- \phi_k\right)\Big]\int\!\!\Big[\prod_{1,2,3} dx_i dy_i dw_i \Big] dw_0\ |\widetilde{\gamma}|\
\Psi'^* O\Psi.
\label{measure}
\end{equation}
The explicit form of the intrinsic $\widetilde{\gamma}$,
\begin{eqnarray}
&& \! \!\! \! \!\! \!\! \! \! \!\! \! \!\! \!\! \! \!\! \!\widetilde{\gamma}=\! -{1\over 4}\!\left(\begin{array}{c | c }
  \begin{array}{c c c  } \! \!\! \! \!\!
 \phi_2 + \phi_3 &0&0 \\ 0& \! \!\! \! \!\! \phi_3 + \phi_1 &0 \\ 0 &0 & \! \!\! \! \!\!  \phi_1 + \phi_2  \! \!\!                                                               
 \end{array}                &-2\overline{S}^{T} \!\!\left(-{3\over 2}v,w\right) \\
 \hline
 -2\overline{S}\left(-{3\over 2}v,w\right)                 &\begin{array}{c c c| c c } \! \!\!
4\phi_1+\phi_2 + \phi_3 &0&0&0&0 \\ 0& \! \!\! \! \!\! \phi_1+4\phi_2 + \phi_3&0&0&0 \\ 0 &0 & \! \!\! \! \!\! \phi_1+\phi_2 +4 \phi_3&0&0 \\ \hline
 0&0&0&\phi_1+\phi_2 +4 \phi_3&-\sqrt{3}(\phi_1-\phi_2) \! \!\! \! \!\! \\ 0&0&0&-\sqrt{3}(\phi_1-\phi_2)&3(\phi_1+\phi_2)  \! \!\! \end{array}
\end{array}\! \!\right),
\end{eqnarray}
shows that, in contrast to the $SU(2)$ case, transition to the intrinsic system does not completely diagonalize $\gamma$.
The determinant $|\widetilde{\gamma}|$ and the inverse $\widetilde{\gamma}^{-1}$, appearing in the Hamiltonian (\ref{uncYMHin}) and the measure
(\ref{measure}), are still complicated. In Section 3.4  we shall indicate, how this difficulty could be overcome using a further algebraic transformation,
 in analogy to the $SU(2)$ case.

By inspection of the magnetic potential 
\begin{equation}
V_{magn}(\widehat{S}^{\rm intr})={g^2\over 2}\left[\sum_{\alpha}\!\!\left(\!\widehat{ B}^{\rm intr}_{\alpha}\!(\widehat{S}^{\rm intr})\!\right)^{\! 2}
+\sum_{i}\left({\cal B}^{\rm intr}_i\!(v,w)\!\right)^{\! 2}
+{3\over 2}\sum_{A}\!\!\left(\!\widetilde{B}_{A}^{(2)}\!(\widehat{S}^{\rm intr})\!\right)^{\! 2}\right],
\end{equation}
with the "symmetric" components $\widehat{B}^{\rm intr}_{\alpha}$  and ${\cal B}^{\rm intr}_{i}$  
\begin{eqnarray}
&& B_{\phi_1}^{\rm intr}={1\over 2} \phi _{2} \phi _{3}+{1\over 4}x_1^2-{1\over 6}(w_0^2+w_1^2+w_2^2+w_3^2)
 +{1\over 2}\!\left(\! w_2\!\left({1\over \sqrt{6}} x_2+y_2\!\right)\!-\!w_3\!\left({1\over \sqrt{6}} x_3+y_3\!\right)\!\! \right)\!
 -\!{1\over 2}\sqrt{3\over 2}\left(x_2 y_2+x_3 y_3\right),\nonumber\\
&&B^{\rm intr}_{x_1}={1\over 2}\phi_{1}x_1-{1\over 2 }\sqrt{3\over 2}(\phi_{2}+\phi_{3})y_1
-{1\over 2\sqrt{6}}\left(\phi_{2}-\phi_{3}\right)w_1
,\quad\quad 
B^{\rm intr}_{y_1}=-{1\over 2 }\sqrt{3\over 2}(\phi_{2}+\phi_{3})x_1-{1\over 2}\left(\phi_{2}-\phi_{3}\right)w_1,
\nonumber\\
&&B^{\rm intr}_{w_1}=-{1\over 2}(\phi_{2}-\phi_{3})\left({1\over  \sqrt{6}}x_1+y_1\right)
-{1\over 3}(\phi_{1}+\phi_{2}+\phi_{3}) w_1,
\quad\quad B^{\rm intr}_{ w_0}=-{1\over 3}(\phi_{1}+\phi_{2}+\phi_{3})w_0,
\nonumber\\
&&{\cal B}^{\rm intr}_{1}=
 -{1\over 2\sqrt{2}}x_2x_3 -{1\over 2\sqrt{3}}( x_2 w_3- x_3 w _2)-{1\over \sqrt{2}}w_2 w_3+{1\over \sqrt{2}}w_0\left({1\over \sqrt{6}}x_1+y_1\right),\end{eqnarray}
(and the corresponding cyclic permutations) and the non-vanishing "antisymmetric" parts  
$\widetilde{B}_{A^\prime}^{(2)}$
\begin{eqnarray}
&&\widetilde{B}^{(2)}_{1+}=-{1\over 3 \sqrt{2}}(\phi_{2}-\phi_{3})\left(x_1- \sqrt{3\over 2}y_1\right)+{1\over 3\sqrt{3}}\left(\phi_{1}-{1\over 2}\phi_{2}-{1\over 2}\phi_{3}\right) w_1~,\quad\quad
\widetilde{B}^{(2)} _{1-},\widetilde{B}^{(2)} _{2-} \ {\rm cycl.\ perm.}\nonumber\\
&&\widetilde{B}^{(2)}_{2+}={1\over 3\sqrt{3}}\left(\phi_{3}-{1\over 2} \phi_{1}-{1\over 2} \phi_{2}\right)w_0~,
\quad\quad
\widetilde{B}^{(2)}_{0}={1\over 6}\left( \phi_{1}- \phi_{2}\right)w_0~,
\end{eqnarray}
one finds that the magnetic potential has the zero-energy valleys 
\begin{equation}
\label{toron}
\left(\widehat{S}_{B^2=0}^{\rm intr}(\phi_3,y_3)\ \ :\ \ \phi_3\ {\rm and}\  y_3\  {\rm  arbitrary }\quad \wedge \quad  {\rm all\ others\ zero}\right) ~.
\end{equation}
Its cyclic permutations are also zero-energy valleys but are excluded by the ordering of the eigenvalues (\ref{range}).
They corresponding valleys $A_{B^2=0}$ in \cite{Luescher} are related to (\ref{toron}) via the special (\ref{eq:pcantr}) transformation
\begin{equation}
\label{unctoron}
A_{B^2=0}(\phi_3,y_3) \equiv \phi_3{\tau_7\over 2}+y_3{\tau_8\over 2}=
 R^{\rm (ad)}(0,0,\pi/4)\ O^\prime(\pi/4,0)\ \widehat{S}_{B^2=0}^{\rm intr}(\phi_3,y_3)~,
\end{equation}
using the $SU(3)$ Euler representation (\ref{adjoint SU3}).
These classical zero-energy valleys of the chromomagnetic 
potential have very important consequences for the quantum level. Since the valleys are
narrowing down with increasing fields, they lead to a discrete quantum spectrum \cite{Luescher,Weisz}, although classically they extend to arbitrarily
large field values. Furthermore, analogous to the $SU(2)$ case of one large field \cite{pavel1}, as a relict of the infinite length of the valleys, 
all excitation energy is expected to go into the increase of the expectation values of these two fields, whereas the expectation values of all the
other fields should remain at there vacuum values required to satisfy the uncertainty relations.


\subsection{\normalsize\bf Symmetries of the unconstrained Hamiltonian}

As a relic of the rotational invariance of the initial gauge field theory
the Hamiltonian (\ref{uncYMHin}) possesses the symmetry
\begin{equation}
\label{cHJ}
[H,J_k]=0~,
\end{equation}
with the total angular momentum operators $J_i= R_{ij}\xi_j$ in (\ref{uncJin}), satisfying
$[J_i, J_j] = i \epsilon_{ijk} J_k$ and  $[J_i, \xi_j] = 0~$.
Hence the eigenstates of $H$ can be characterized by the quantum numbers $J$ and $M$.
Furthermore $H$ is invariant under arbitrary permutations $\sigma_{ij}$ of any two of the
three indices $1,2,3$,
time reflections T, and parity reflections $ P: \phi_i\rightarrow -\phi_i \wedge \overline{S}\rightarrow - \overline{S}$, and 
charge conjugation $ C: \phi_i\rightarrow \phi_i  \wedge \overline{S}\rightarrow - \overline{S}$
\begin{equation}
\label{cHsTPC}
[H,\sigma_{ij}]=0~,\ \ \ \ \ \ [H,T]=0~,\ \ \ \ \ \ [H,P]=0~,\ \ \ \ \ \ [H,C]=0~.
\end{equation}
As a consequence, the energy eigenfunctions  can be chosen real, and invariance of a functional under $P$ and $C$ implies 
that it contains an even number of factors $\in \{\phi_i\}$ and  an even number of factors $\in \{x_i,y_i,w_i,w_0\} $ .

\subsection{\normalsize\bf Virial theorem}

Writing $H=\frac{1}{2}\left(E^2+B^2\right)$ and denoting
the eigenstates of H by $|n\rangle $, with energies $E_n$,
one obtains the virial theorem
\begin{equation}
\label{VirialTh}
\langle n| E^2|n \rangle = 2 \langle n| B^2|n \rangle~,
\end{equation}
a very useful tool to judge
the quality of the approximate eigenstates obtained using the
variational approach.

\subsection{\large\bf Outlook: Towards a calculation of the low-energy eigensystem}

As in the $SU(2)$ case, a variational calculation should be carried out, 
under full consideration of all  symmetries (\ref{cHJ}) and  (\ref{cHsTPC}) and the Virial theorem (\ref{VirialTh}),
with basis functions from the corresponding harmonic oscillator problem, replacing the magnetic potential by higher dimensional
harmonic oscillator potential.
In the  first place the spin-0 sector ${\bf J}^2=\xi^2=0$ of the Hamiltonian (\ref{uncYMHin}) should be investigated, where the eigenstates 
are cyclic singlets. 
We point out that the $SU(3)$ energy spectrum of the original constrained system (\ref{G_a}) has already been studied
in \cite{Weisz} using the variational approach. Numerically constructing step by step a low-energy basis of gauge- and rotation-invariant polynomials up to a certain degree (10), 
a rather good value for the ground state and less accurate values for the lowest excitations have been obtained. 

The unconstrained approach presented in this work, where gauge invariance is already manifest after the implementation of the Gauss laws, 
should reproduce these results in a very effective and accurate way.
 The main obstacle towards a calculation of the eigensystem of  the Hamiltonian (\ref{uncYMHin}) is, that 
the FP determinant $|\widetilde{\gamma}|$ is rather complicated. 
In contrast to the $SU(2)$ case, through the transition to intrinsic coordinates, the FP matrix  is not completely diagonalized for the $SU(3)$ case.
To make progress, a further algebraic transformation has to be carried out, similar to the transition from $\{\phi_1,\phi_2,\phi_3\}$ to
the corresponding elementary symmetric polynomials in the $SU(2)$ case \cite{pavel1}, which cancels the non-trivial Jacobian. 
Since such an algebraic transformation is equivalent to a transition to the spin-0 combinations
$\{S^{(0)},(SS)^{(0)},(SSS)^{(0)}\}$ using the spin-0 and spin-2 components of the symmetric tensor field $S$, 
we have to look here  for the corresponding transformation to spin-0 combinations of spin-0, spin-1, spin-2, and spin-3.
In particular the FP determinant has to be expressed in terms of the new algebraic variables,
in order to find its zeros, to limit the range of integration of the field variables.
Preliminary investigations, including only 6 degrees of freedom, the $S$ and the vector field $V^{(1)}$, a bijective transformation
from $\{ \phi_1,\phi_2,\phi_3,v^{(1)}_1,v^{(1)}_2,v^{(1)}_3\}$ to the set of  spin-0 combinations $\{S^{(0)},(SS)^{(0)},(SSS)^{(0)},
(VV)^{(0)},(SVV)^{(0)},(SSVV)^{(0)}\}$ can be constructed. The generalization to include also the spin-3 field $W^{(3)}$ is more difficult
and under current investigation.

\section{\large\bf Conclusions}

It has been shown in this work how an unconstrained Hamiltonian formulation of $SU(3)$ Yang-Mills quantum mechanics of spatially constant fields can be obtained using the method of minimal embedding of $SU(2)$ into $SU(3)$ by Kihlberg and Marnelius \cite{Kihlberg and Marnelius}. This has lead us to
a novel polar decomposition (\ref{eq:pcantr}) of the spatial components of the gauge fields, in terms of which the non-Abelian Gauss laws can be Abelianized and be implemented exactly, to obtain an cyclic symmetric unconstrained Hamiltonian of colorless spin-0,spin-1,spin-2, and spin-3 dynamical glueball degrees of freedom.
Furthermore, a transformation of the 16  physical variables to 3 rotational degrees of freedom and 13 invariants under spatial rotations has been carried out.
It has been shown that the chromomagnetic potential has classical zero-energy valleys
for two arbitrarily large classical glueball fields, which are the unconstrained analogs 
of the well-known constant Abelian fields. On the quantum level, 
all glueball excitation energy is expected \cite{pavel1} to go into the increase of strengths of these two fields.

Finally, an outview has been given, how the eigensytem of the obtained unconstrained $SU(3)$ Hamiltonian could be obtained in a similarly effective and 
accurate way as for the $SU(2)$ case \cite{pavel1}, as an alternative to the already existing $SU(3)$ calculation by Weisz and Zimann \cite{Weisz}.
The main problem in the unconstrained approach proposed in the present work is, that the FP determinant, even after
transition to the intrinsic system, is still quite complicated. This has to be overcome by using a further algebraic transformation
to new variables in generalization of the transition to elementary symmetric polynomials used for the $SU(2)$ case.

We mention, that  in the work  by Dahmen and Raabe \cite{Dahmen and Raabe} almost 20 years ago, considering the classical $SU(3)$ Yang-Mills 
mechanics for total spin-0, the minimal embedding of $SU(2)$ into $SU(3)$ has already been used 
to implement a condition similar to the symmetric gauge condition (\ref{symgauge}).
The present work goes well beyond their work in several respects: Firstly, we have carried out a separation of the gauge and rotational degrees of freedom and have obtained the unconstrained Hamiltonian for arbitrary total spin. Using the novel polar decomposition (\ref{eq:pcantr}), we have first carried out
an exact gauge reduction and only afterwards we have applied another coordinate transformation of the unconstrained system to
rotational and rotation invariant variables.
Secondly, the unconstrained Hamiltonian obtained here, has explicit cyclic symmetry in the indices 1,2,3, apart from rotational symmetry, parity, and charge conjugation symmetry. Thirdly, the reduction is carried out on the quantum level and will allow for a comparison with existing investigations on the quantum level, such as that by Weisz and Zimann \cite{Weisz} .

The generalization of the approach to field theory is straightforward along the lines of the $SU(2)$ case in \cite{KP1,pavel2},
extending the coordinate transformation (\ref{eq:pcantr}) to
\begin{equation}
\label{coordtrafo}
A_{ak} \left(q, \widehat{S} \right) =
O_{a\hat{a}}\left(q\right) \widehat{S}_{\hat{a}k}
- {1\over 2g}c_{abc} \left( O\left(q\right)
\partial_k O^T\left(q\right)\right)_{bc}~.
\end{equation}
The FP operator of the symmetric gauge (\ref{symgauge}),
\begin{equation}
\label{FP}
^{\ast}\! D_{\hat{a}\hat{b}}( \widehat{S})\equiv c_{\hat{a}\hat{c}i}D_i( \widehat{S})_{\hat{b}\hat{c}}=c_{\hat{a}\hat{c}i}\left(\delta_{\hat{b}\hat{c}}\partial_i-g c_{\hat{b}\hat{c}\hat{d}} \widehat{S}_{\hat{d}i}\right)
=g\gamma_{\hat{a}\hat{b}}+c_{\hat{a}\hat{b}i}\partial_i=
g\gamma_{\hat{a}\hat{a}^\prime}\left(\delta_{\hat{a}^\prime\hat{b}}
+\gamma^{-1}_{\hat{a}\hat{a}^\prime}c_{\hat{a}^\prime\hat{b} i} {1\over g}\partial_i\right)~,
\end{equation}
and hence the corresponding unconstrained Hamiltonian of $SU(3)$ Yang-Mills theory, can be expanded in the number of spatial derivatives in the low energy region, equivalent to a strong coupling expansion, with the Yang-Mills quantum mechanics of spatially constant fields constituting the leading order.

\section*{\large\bf Acknowledgments}

I would like to thank A. Dorokhov and J. Wambach for their interest and support.
Financial support by the LOEWE program HIC for FAIR and in part by the BMBF grant 06-DA-9047-I  are gratefully acknowledged.

\begin{appendix}

\section*{Appendix: Explicit form of the  the intrinsic $\widetilde{T}$}

The spin-1 components $\widetilde{T}^{(1)}_i$ and the spin-2 components $\widetilde{T}^{(2)}_A$ 
of the intrinsic $\widetilde{T}$ read
\begin{eqnarray}
\widetilde{T}^{(1)}_i&=&{1\over 4} \left[\xi_i-2\widetilde{J}_i^{(v)}-{1\over 3}\widetilde{J}_i^{(w)}\right]~,\quad i=1,2,3
\nonumber
\end{eqnarray}
\begin{eqnarray}
\label{uncTin}
\widetilde{T}^{(2)}_{1+}\!\!\!\!&=&\!\!\!\!-{i\over 2 \sqrt{2}}\Bigg[x_1
\left({\partial \over\partial \phi_{2}}-{\partial \over\partial \phi_{3}}\right)-
( \phi_{2}- \phi_{3}){\partial \over\partial x_1}\Bigg]
\!\! -{i\over \sqrt{3}}\Bigg[w_1\!\left({\partial \over\partial \phi_{1}}-{1\over 2}\!\left({\partial \over\partial \phi_{2}}+{\partial \over\partial \phi_{3}}\right)\!\right)\!-\!
\left( \phi_{1}-{1\over 2}\left( \phi_{2}+ \phi_{3}\right)\!\right){\partial \over\partial w_1}\Bigg]\nonumber\\
&&-{1\over 8}\Bigg[\left(x_2
-\sqrt{3\over 2} y_2\right)\widetilde{\cal P}_3 -\left(x_3
-\sqrt{3\over 2} y_3\right)\widetilde{\cal P}_2 \Bigg]~,
\quad\quad\quad\quad
\widetilde{ T}^{(2)}_{1-}~, 
\widetilde{ T}^{(2)}_{2-} \quad {\rm cycl.\ perm.}
\nonumber\\
\widetilde{T }^{(2)}_{2+}\!\!\!\!&=&\!\!\!\!-{i\over \sqrt{3}}\Bigg[w_0\!\left(\!{\partial \over\partial \phi_{3}}\!
-\!{1\over 2}\!\left(\!{\partial \over\partial \phi_{1}}\!+\!{\partial \over\partial \phi_{2}}\right)\!\!\right)\!-\!
\left(\!\phi _{3}\!-\!{1\over 2}\!\left(\phi _{1}\!+\!\phi _{2}\right)\!\right)\!\!{\partial \over\partial w_0}\! \Bigg]
\!\!-\!{1\over 2}\!\left(\!x_3\widetilde{\cal P}_3
-{1\over 2}(x_1\widetilde{\cal P}_1\!+\! x_2\widetilde{\cal P}_2)\!\right)
\! -\!{1\over 2}\sqrt{3\over 2}\!\left(w_1\widetilde{\cal P}_1\!-\!w_2\widetilde{\cal P}_2\right),
\nonumber\\
\widetilde{T}^{(2)}_{0}\!\!\!\!&=&\!\!\!\!-{i\over 2}\Bigg[w_0\left({\partial \over\partial \phi_{1}}-{\partial \over\partial \phi_{2}}\right)-
\left(\phi _{1}-\phi _{2}\right){\partial \over\partial w_0} \Bigg]
\!-\!{\sqrt{3}\over 4}\left(
x_1\widetilde{\cal P}_1
\!\!-x_2\widetilde{\cal P}_2\right)
+{1\over \sqrt{2}}\left(w_3\widetilde{\cal P}_3
\!\!-{1\over 2}w_1\widetilde{\cal P}_1
\!\!-{1\over 2}w_2\widetilde{\cal P}_2\right)~.
\nonumber
\end{eqnarray}
using the abbreviation
$
\widetilde{\cal P}_1:=  \left(\xi_1-\widetilde{J}^{(v)}_1-\widetilde{J}^{(w)}_1\right)/\left(\sqrt{2}(\phi_2 - \phi_3)\right)
$
and its cyclic permutations.
\end{appendix}


\begin{thebibliography}{99}
%
\bibitem{Christ and Lee}
N.H. Christ and T.D. Lee, Phys. Rev. D 22 (1980) 939.
\bibitem{GKMP}
S.A. Gogilidze, A.M. Khvedelidze, D. M. Mladenov and H.-P. Pavel,
Phys. Rev. D 57 (1998) 7488.
%
\bibitem{KP1}
A.M. Khvedelidze and H.-P. Pavel,
Phys. Rev. D 59 (1999) 105017.
%
\bibitem{KMPR}
A.M. Khvedelidze, D. M. Mladenov, H.-P. Pavel, and G. R\"opke,
Phys. Rev. D 67 (2003) 105013.
%
\bibitem{pavel2}
H.-P. Pavel, Phys. Lett. B 685 (2010) 353.
\bibitem{Luescher}
M. L\"uscher , Nucl. Phys. B219 (1983) 233.
%
\bibitem{Luescher and Muenster}
M. L\"uscher and G. M\"unster, Nucl. Phys. B232 (1984) 445.
%
\bibitem{Savvidy}
G. K. Savvidy,, Nucl. Phys. B246 (1984) 302; Phys. Lett. 159B (1985) 325.
%
\bibitem{Medvedev}
B.V. Medvedev,  Theor. Math. Phys.  60 (1984) 224.
%
\bibitem{Simon}
Yu. Simonov, Sov. J. Nucl. Phys. 41 (1985) 835.
%
\bibitem{Koller and van Baal}
J. Koller and P. van Baal, Nucl. Phys. B273 (1986) 387;
Nucl. Phys. B302 (1988) 1;
P. van Baal and J. Koller,
Ann. of Phys. (N.Y.) 174 (1987) 299.
%
\bibitem{KP2}
A. Khvedelidze and H.-P. Pavel,
Phys. Lett. A 267 (2000) 96;
%
A. Khvedelidze, H.-P. Pavel, and  G. R\"opke,
Phys. Rev. D 61 (2000) 025017.
%
\bibitem{pavel1}
H.-P. Pavel, Phys. Lett. B 648 (2007) 97.
%
\bibitem{pavel3}
H.-P. Pavel, Phys. Lett. B 700 (2011) 265.
\bibitem{Weisz}
P. Weisz and V. Ziemann, Nucl. Phys. B284 (1987) 157.
%
\bibitem{Dahmen and Raabe}
B. Dahmen and B. Raabe, Nucl. Phys. B384 (1992) 352.
%
\bibitem{Gerdt2}
V.P. Gerdt, Y.G. Palii, and A.M. Khvedelidze, Theor. Math. Phys.  155(1) (2008) 557.
\bibitem{Kihlberg and Marnelius}
A. Kihlberg and R. Marnelius,
Phys. Rev. D 26 (1982) 2003.
%
\bibitem{Cornwell}
J.F. Cornwell, Group theory in physics, Vol. II (Academic Press, London, 1984).
%
\bibitem{Olive}
D. Olive, Proc. Int. Summer Inst. on Theor. Physics, Bad Honnef 1980 (Plenum Press, New York, 1981), p. 207.
\bibitem{Byrd}
M. Byrd, "The geometry of SU(3)" (1997) {\it Preprint} quant-ph/9708015; 
 M. Byrd, J. Math. Phys. {\bf 39} (1998) 6125.
%
\bibitem{Tilma}
T. Tilma and E.C.G. Sudarshan, J. Phys. A {\bf 35} (2002) 10467.
%
\bibitem{Gerdt}
V. Gerdt et al., J. Math. Phys. {\bf 47} (2006) 112902.
%


\end{thebibliography}
\end{document}